\documentclass[11pt,a4paper]{article}

\usepackage[utf8]{inputenc}
\usepackage{amsmath}
\usepackage{amsfonts}
\usepackage{amssymb}
\usepackage{lmodern}
\usepackage[left=2.5cm,right=2.5cm,top=2.5cm,bottom=2.5cm]{geometry}
\usepackage{tensor}
\usepackage{braket}
\usepackage[nottoc]{tocbibind} 
\usepackage{bbm} 
\usepackage{slashed} 
\usepackage{color}

\usepackage{color} 
\usepackage{hyperref} 
\hypersetup{
    colorlinks=true, 
    linkcolor= black, 
    urlcolor= black, 
    citecolor = black,
    linktoc=all 
}

\allowdisplaybreaks  

\numberwithin{equation}{section} 

\DeclareMathOperator{\tr}{tr}

\newcommand{\dd}[1]{d#1\ }
\DeclareMathOperator{\Tr}{Tr}

\newcommand{\su}{ { \text{susy} } }
\newcommand{\Lagr}{ { \mathcal{L} } }

\newcommand{\SYM}{ { \text{SYM} } }

\newcommand{\covD}{ { \mathcal{\nabla} } }

 \def \b {b} \def \ed {\end{document}}
 \def \iffa {\iffalse} 
\def \ci {\cite} 
\def\foot{\footnote}
\newcommand{\rf}[1]{(\ref{#1})}
\def\la{\label}
\def \ov {\over} 

\def \g {\gamma} \def \b {\beta}
\def \no {\nonumber}
 \def \be {\begin{equation}}\def  \ee {\end{equation}} \def \ha {{1\ov 2}}
\def \WW {F}\def \su {{\rm susy}}
\def \tA  {\tilde A}\def \rr {\ell} 
\def \su {{(1,0)}}

\def \k {\kappa}

\begin{document}

\begin{flushright}\small{Imperial-TP-AT-2019-{04}}

\end{flushright}
\vspace{2.5cm}
\begin{center}

{\Large\bf  One-loop $\b$-functions  \\ \vspace{0.2cm}
 in 4-derivative  gauge theory in 6  dimensions 
}

\vspace{1.2cm}

{Lorenzo Casarin$^{a,}$\footnote{ lorenzo.casarin@aei.mpg.de } and
Arkady A. Tseytlin$^{b,c,}$\footnote{ Also at the   
Lebedev Institute, Moscow.
\ \  tseytlin@imperial.ac.uk }
}

\vspace{0.5cm}

{\em
\vspace{0.15cm}
$^{a}$ Max-Planck-Institut f\"u{}r Gravitationsphysik (Albert-Einstein-Institut)  \\
\vspace{0.05cm}
 Am M\"u{}hlenberg 1, DE-14476 Potsdam, Germany
\\
\vspace{0.15cm}
$^{b}$ Blackett Laboratory, Imperial College, London SW7 2AZ, U.K.\\
\vspace{0.15cm}
$^{c}$ Institute of Theoretical and Mathematical Physics\\
\vspace{0.05cm}
 Moscow State University,  119991, Russia }

\end{center}
\vspace{0.9cm}
%
\begin{abstract}
\noindent 
A  classically scale-invariant 6d  analog  of  the  4d Yang-Mills  theory   is the 
4-derivative $     (\nabla F)^2 +    F^3$  gauge theory  with two independent couplings.  
Motivated by  a search  for  a perturbatively conformal but possibly non-unitary
6d models    we compute the  one-loop  $\b$-functions  in this theory. 
A systematic way of doing this using the  background field method 
 requires  the (previously unknown) expression for the   $b_6$ Seeley-DeWitt
coefficient  for  a  generic  4-derivative  operator; we derive it here. 
As an application, we  also  compute  the one-loop  $\b$-function 
 in the (1,0)   supersymmetric $ (\nabla F)^2$ 
6d gauge   theory   constructed  in hep-th/0505082.
\end{abstract}

\date{\today}  \date{\currenttime}

\newpage
\tableofcontents

\setcounter{footnote}{0}

\section{Introduction}

Like Einstein theory in 4 dimensions,  the  6d Yang-Mills theory   with  the standard $F^2$ action  
has dimensional coupling   and is not power-counting renormalizable.
 A 6d analog of the classically scale invariant and 
renormalizable  $R^2 +  C^2$  4d gravity  is the 4-derivative 
$   (\nabla F)^2 +     F^3$  gauge theory. 
Such 4-derivative  
terms  are  induced as counterterms  
when considering the standard 
scalars,  fermions   or YM vectors coupled to a  background gauge field  in 6d \ci{Fradkin:1982kf}.
While non-unitary, this model   may serve as  a   building block  of  possible  higher-derivative 
  (super)conformal theories  in 6 dimensions.\footnote{ In 4 dimensions 
the $  F^2 +    (\nabla F)^2 +      F^3$  theory 
was  studied   in \ci{Fradkin:1981iu}   and  later   in \ci{Grinstein:2008qq}.
The result of  \ci{Fradkin:1981iu} for the   one-loop  divergences in this 4d theory was corrected in \ci{Casarin:2017xez}
making it in  agreement with that of \ci{Grinstein:2008qq}.
 }
Similar 4-derivative  6d   gauge theories were discussed, e.g.,  in  
 \ci{Ivanov:2005qf,Smilga:2006ax,Beccaria:2015uta,Beccaria:2015ypa,Giombi:2015haa,Giombi:2016fct,
 Osborn:2016bev,Huang:2018hho,Johansson:2018ues}. 

The aim of the present paper is to compute the one-loop  $\b$-functions   in  the  
 Euclidean 
6d theory 
with the action\foot{ 
 We use $m,n,k, ...= 1, ..., 6$ for coordinate indices  and  flat  Euclidean 6d metric so that 
   the position of contracted indices is irrelevant. 
   The gauge group generators are normalized as 
$\tr( t^a t^b) =-  T_R \delta^{ab}, \ [t^a, t^b]= f^{abc} t^c$,   where $T_R= \ha$ in the fundamental representation of $SU(N)$ 
(we denote the trace in this case as \(\Tr{}\))  and $T_R= C_2 = N $ in the  adjoint representation.
  }
\begin{equation}
\label{zaa}
\begin{split}
S & = - \frac{1}{g^2}\int d^6 x \,  \Tr\Big[
		\left(\covD_m {F}_{mn}\right)^2
		+ 2 \gamma F_{mn} F_{nk} F_{km} 
	\Big]\\
&=
\frac{1}{2g^2}\int d^6x \,  \Big[
		\left(\covD_m F^a_{mn}\right)^2
		+ \gamma f^{abc} F^a_{mn} F^b_{nk} F^c_{km}
	\Big]\ . 
\end{split}
\end{equation}
Here  
 $g$ and $\g$ 
are  the  two independent 
dimensionless  coupling parameters.\foot{ Two other  possible 
4-derivative $ \nabla F \nabla F $   invariants are  related to the above two by the Bianchi identity, e.g., 
$\ {F}_{mn}\covD^2 {F}_{mn} = -2  \left(\covD_m {F}_{mn}\right)^2   + 4 {F}_{mn} {F}_{nk} {F}_{km} + \text{total derivative}$.}

In general, the UV  logarithmically divergent part of the  6d one-loop effective action \( \Gamma_{1} \) in a  
gauge field background 
may  be written as\foot{ Here $\tr$  is the trace over the matrix indices  of a particular representation to which  the quantum field belongs; 
for example, in the gauge theory  case it is in the adjoint representation 
$A^{ab}_m = f^{a c b } A^c_m, \ f_{acd} f_{bcd} = C_2 \delta_{ab}$.} 
\begin{equation}\label{zba}
\Gamma_{1\infty } = -  \frac{\log \Lambda}{(4\pi)^3} \;
	\int \dd{^6x}  \tr 
		 \Big[
			- \frac{1}{60} \b_2 \left(\nabla_m F_{mn}\right)^2
			+ \frac{1}{90} \b_3 F_{mn} F_{nk} F_{km}
		\Big]\ , 
\end{equation}
where 
the 1-loop $\b$-function coefficients  $\b_{2}$, $\b_3$ depend on the field content of the theory. 
As we shall  find below, their values in the case of the 4-derivative theory \rf{zaa} are given by 
the following functions of  the  coupling  $\g$ 
(1-loop coefficients do not depend on the overall $g^2$ coupling)
\begin{align}\label{zbb}
\b_{ 2A }  & =   249
	\;   ,
&
\b_{ 3A } & =     9 - 900 \gamma + \frac{405}{2}  \gamma^3
	\;   .  
\end{align}
Somewhat surprisingly,  the coefficient  $\b_{2A} $   of the $(\nabla F)^2$ 
 divergence   turns out to be  independent of the coupling $\gamma$.
 
The total values of $\b_2$, $\b_3$   in a 6d renormalizable model 
  containing   the gauge theory \rf{zaa}  minimally coupled to 
the ordinary-derivative ``matter" fields --
$N_0$ real scalars, $N_{\frac12}$ Weyl fermions,  
 \(N_1\)    YM vectors  and  $N_T$ self-dual tensors (interacting with   $A_m$   as  in  \ci{Huang:2018hho}) 
 are then   \ci{Fradkin:1982kf,Huang:2018hho}\foot{ Here  all the fields are taken for simplicity in the adjoint representation; 
 in the case of other representations  one is to rescale the numbers $N_s$  by $T_R/C_2$. 
 We  corrected misprints in \ci{Fradkin:1982kf} mentioned in \ci{Huang:2018hho}.
 Note that the  vector $N_1$ terms   here are formal: they  indicate  the 6d  YM contribution  in the absence of higher-derivative terms in \rf{zaa}. 
 In the combined $F^2 + (\nabla F)^2 + F^3$ theory  discussed below in Appendix~\ref{C}
 the values of $\b_{ 2 }$  and $\b_3$  are the same as in  the theory \rf{zaa} without the YM  $F^2$ term. }
 \begin{equation}\label{zbc}
\begin{aligned}
\b_{ 2 } & =   \b_{ 2A }  \,  -27\, N_T  -  36\, N_1 + N_0  + 16\, N_{\frac12} \; ,
\\*
\b_{ 3 } & =  \b_{ 3A }\,   - 57\, N_T  +   4\, N_1 + N_0 - 4\, N_{\frac12}  \ . 
\end{aligned}
\end{equation}
Note that for  the ordinary spin 0, 1/2, 1  fields  their  contributions to $\b_3$ are proportional to 
the number of dynamical  degrees of freedom. The same  is true  also for the   4-derivative  gauge theory  \rf{zaa} 
with $\g=0$: $ \b_{ 3A }=9$   is the  number of d.o.f.\ of a 
4-derivative gauge vector   in 6d.\foot{ While the  2-derivative YM   vector in $d$ dimensions 
   has $\ha 2 ( d-1) -1 = d-2$  dynamical d.o.f., 
for the 4-derivative gauge vector in  \rf{zaa}  one finds $\ha 4 ( d-1) -1 =2d-3$, i.e.  5 in $d=4$ and  9    in $d=6$.} 
  As a consequence one should   get  $\b_3=0$  in a supersymmetric  theory; this  is consistent with the non-existence 
of a super-invariant containing $\tr (F_{mn} F_{nk} F_{km}) $. 
Indeed,  for  the standard   2-derivative 6d (1,0) SYM  theory ($N_1=1,\ N_{\frac12}=1$)     and  for the scalar (hyper)
multiplet  ($N_0=4,\ N_{\frac12}=1$) one finds 
\be \la{15}  \b_{2\, _{(1,0)\, \SYM}} = -20 \ , \qquad \qquad  \b_{2\,\rm  scal} = 20 \ , \qquad \qquad \b_{3\,\rm _{(1,0)\, \SYM}} =\b_{3\,\rm scal} =0  \ . \ee
Since $\nabla_m F_{mn}=0$ on the standard YM equations of motion  the (1,0) SYM  theory is 1-loop finite on shell. 
The sum  of the contributions of the  two  multiplets in \rf{15}
  corresponds to  the (1,1)  SYM theory in 6d  (and thus   to $N=4$ SYM in 4d)  which is 
   1-loop finite even off-shell   \ci{Fradkin:1982kf}
\be  \la{155}  \b_{2\, _{(1,1)\, \SYM}} = \b_{3\,\rm _{(1,1)\, \SYM}}  =0  \ . \ee
In the     $(1,0)$ supersymmetric 4-derivative gauge   theory with the action given by the super-extension  \cite{Ivanov:2005qf} of 
$ \tr	\left(\covD_m {F}_{mn}\right)^2$
 (containing also  interacting $\slashed{\nabla}^3 $ Weyl  fermion  and  three 
$\nabla^2$ scalars) 
 we  will  find below  that 
\be\label{zbd}
\b_{ 2\, \su }  =  220\;  , \qquad \qquad  \b_{ 3\, \su }  =  0 \,  . 
\ee
This  result  is in agreement (modulo  notation change) with the one 
 given   in  the recently revised    version  of \cite{Ivanov:2005qf}. 
This  theory is   non-unitary  and is  also formally  inconsistent  having
 a chiral anomaly   \ci{Smilga:2006ax} (the same as in the (1,0) 6d SYM theory  containing Weyl fermion). 
One may still  hope to cancel  all of its  anomalies by adding some higher derivative  6d ``matter''  multiplets (cf.
 \ci{Ivanov:2005kz,Kuzenko:2017jdy,Kuzenko:2017xgh}).


The calculation of the  $\beta$-functions \rf{zbb} 
is most  straightforward in  the  background field method   and using the heat kernel expansion to extract the log divergences 
of the determinants. 
This 
 requires the knowledge of the corresponding  $b_6$  Seeley-DeWitt coefficient  for the {  4-derivative} operator $\Delta_4= \nabla^4 + ...$ 
 in a gauge field background.  While  \( b_6\) is  available   for the 2-derivative $\Delta_2$ 
 operators \ci{Gilkey:1975iq},  its  expression for
 $\Delta_4$   was not known so far. 
 The main new   technical result  of this paper is  the computation 
    of $b_6 (\Delta_4)$. 
    We shall      use the same  strategy as employed  previously in \cite{Fradkin:1981iu}
 to obtain \( b_4(\Delta_4)\) from the known  expression for  \( b_4(\Delta_2)\)
  by  considering  special  factorized  cases of the operator $\Delta_4$.

It would be interesting to extend the computation of the $b_6$ coefficient  for the  4-derivative operators  to the case of a 
 curved metric background (finding the analog of the corresponding expression for $b_4$  in  \cite{Fradkin:1981iu}).
 This  would allow, in particular, to compute  the one-loop   UV divergences    in $d=6$  conformal supergravity 
  and verify the  expectation    \ci{Beccaria:2015uta,Beccaria:2015ypa} 
  that the  higher derivative (2,0) 6d conformal supergravity coupled to exactly 26 (2,0) tensor multiplets has the 
   vanishing  conformal anomaly.\foot{This is the 6d counterpart of the known fact of cancellation of the conformal anomaly in the 4d system of $\cal N$=4 conformal supergravity coupled to 4 vector $\cal N$=4 multiplets \ci{Fradkin:1983tg,Fradkin:1985am}.}
Another important step would be to extend the background field  approach to the computation of   UV divergences 
in 4-derivative gauge   theories to the  two-loop level  generalizing  the  methods of \ci{Jack:1982hf,Jack:1983sk,Grinstein:2015ina}.

The  rest of the  paper is organized as follows.
In section~2  we  present the general form of  the  one-loop effective action of  the  theory \rf{zaa}. 
In section~3 the result  for the   heat kernel    coefficient $b_6$  that controls  the logarithmic  divergence of the 
determinant of    a generic  4-derivative   operator  is    given. 
  In section 4 this expression is   applied to compute the one-loop divergences 
in the bosonic  gauge theory  \rf{zaa}   and its (1,0) supersymmetric extension 
(with $\g=0$). 
Details of the derivation of $b_6(\Delta_4)$ are  described in   Appendix~\ref{B}.
In  Appendix~\ref{C}  we  discuss   divergences of the 
combined 2- and 4-derivative   $ {1\ov g^2} [ \kappa^2 F^2 +    (\nabla F)^2 +  \gamma    F^3]$  gauge theory and its (1,0) supersymmetric version:
adding $F^2$ does not change  the $\b$-functions \rf{zbb} for $g$ and $\g$ but leads to the $\g$-dependent  
 $\b$-function for $\kappa$. 
  
\section{One-loop  effective action }

The   derivation of the one-loop effective  action in the 4-derivative theory \rf{zaa}  in 6d follows the same steps as in the 
4d  case discussed in Appendix C   of    \ci{Fradkin:1981iu} (for a review, see also  \ci{Casarin:2017xez}). 
Expanding the invariants in  \rf{zaa}  near a classical background $A_m^a \to {A}_m^a + \tA_m^a$  we  get  
\begin{align}
\no 
\Tr \left(\covD_m F_{mn}\right)^2 
	& \to
	-  \frac12 \tA_m^a\,  \Big[
			\delta_{mn}\covD^4
			+	4F_{mn} \covD^2
			-
				 2 \left(
					  \covD_k F\indices{_{km}} \delta_{nr} 
					+ 2 \covD_k F\indices{_{k[n}} \delta_{r]m} 
				\right) \covD_r
\\   \label{xab}
& \hspace{6em}
    {}	-2 \covD_n \covD_k F_{km}
			+ 4 F_{mk} F_{kn}  
		\Big]^{a b} \tA_n^b 
			- \frac12 (\covD_m \tA_m^a)\covD^2(\covD_n \tA_n^a)\ , 
\\
\no
\Tr (F_{mn} F_{nk} F_{km})
	& \to
	  \tA_m^a \, \Big[\Big(
			\frac{3}{2} F\indices{_{[m}^{(r}}\delta_{n]}^{k)}
			- \frac{3}{4} F_{mn} \delta^{rk} 
		\Big) \covD_r \covD_k
	+  
			3 \covD_k F\indices{_{[m}^{[r}} \delta_{k]}^{n]} 
		 \covD_r
\\  \label{xac}
& \hspace{6em}
	-   \Big(
			\frac{3}{4} [F_{mk}, F_{kn}]
			+ \frac{3}{4} F_{r(m} F_{n)r}
			+ \frac{3}{8} F_{rk} F_{rk} \delta_{mn} 
		\Big) \Big]^{ab} \tA^b_n \ , 
\end{align}
where $F_{mn}$ and $\nabla_m$ depend on the background $A_m$ and $a,b$ are indices in the adjoint representation. 
Then  the  quadratic  part of the fluctuation Lagrangian in \rf{zaa}  may be written as 
\be \la{xae}
 \Lagr^{(2)}  =
\frac{1}{2g^2} \tA^a_m\, (\Delta_{4A})_{mn}^{ab}\, \tA_n^b
+ \frac{1}{2g^2} (\covD_m \tA^a_m)
	 \big( - \covD^2 
	  \big) (\covD_n \tA^a_n)  \ .  \ee
The  second term  here  can be cancelled by adding a gauge-fixing ($\nabla_m \tilde A_m=f(x)$)  term  averaged with the  operator $-\nabla^2$.
The 4-derivative operator $\Delta_{4A}$  acting on $\tilde A^a_m$  can be written in
 the following ``symmetric"
 form
\begin{align}\label{zadd}
	\Delta_{4}
& =
	 \covD^4
			+ \covD_r  \hat{ V }_{rk}   \covD_k
			+ \hat{N}_k \covD_k
			+ \covD_k  \hat{N}_k 
			+ \hat{U},
& & 
	\hat{ V }_{rk}  = \hat{V}_{kr} \  , 
\end{align}
where 
$  \hat{ V }_{rk} $, $\hat{N}_k $, $\hat{U}$ are local covariant matrices in the internal $(a,m),(b,n)$ indices 
reading
\begin{align}
\nonumber
	(\hat{V}_{rk})_{mn}
& =
	(4+3\gamma) F_{mn}\delta^{rk} 
	-
	6 \gamma F\indices{_{[m}^{(r}}\delta^{k)}_{n]}\ , 
\\
\label{xag}
	(\hat{N}_{k})_{mn} 
& =
	\ha \big( 2+  3  \gamma \big)\covD_r F_{rk} \delta_{mn}
	- \ha \big( 4+  3  \gamma\big) \covD_r F_{r(m} \delta_{n)k}
	- \frac32 \gamma \covD_{(m} F_{n)k}\ , 
\\\nonumber
	(\hat{U})_{mn} 
& = 
	-\ha \big(4+ 3  \gamma\big) F_{kn}F_{mk}\
	+ \frac{3}{2} \big(4+ 3  \gamma\big)F_{km}F_{nk}
	+ \frac32 \gamma F_{rk} F_{rk} \delta_{mn}
	+ 3 \covD^2 F_{mn} \ . 
\end{align}
The operator that appears 
in the effective action after path-integral  is performed  (i.e. $\Delta_{4A}$ in \rf{xae})   should 
 be self-adjoint  and  this is so for \rf{zadd} with   \rf{xag}.\foot{ \la{zae}
Note that \rf{zadd}  is a completely general form for a fourth-order elliptic differential operator without the three-derivative term. The self-adjointness can be imposed via the following additional conditions on the coefficients
 $
\hat{V}_{mn}^{\dagger}  = \hat{V}_{mn}  $,
$\   \hat{N}_m^{\dagger}  =- \hat{N}_m   $,
$ \   \hat{U}^{\dagger}  = \hat{U}$ where $\dagger$ is transposition if the field is real, and hermitian conjugation if the field is complex.}

The  1-loop effective action is then given by 
\begin{equation}\label{zah}
\Gamma_{1} = \frac12 \log\frac{\det \Delta_{4A} }{(\det \Delta_{\rm gh})^2 \,\det H     }=    \frac12 \log{\det \Delta_{4A} }
-  \frac32  \log{\det  \Delta_{2, 0}}\ , \qquad   \Delta_{2, 0}  = - \nabla^2 \ , 
\end{equation}
where $\Delta_{\rm gh} = -\nabla^2$ is the ghost operator and $ H = -\nabla^2  $ is the gauge-condition averaging operator
required to cancel the last term in   \rf{xae}.  Using the proper-time cutoff,   
 the log divergent part of a determinant   can be expressed 
(in general dimension $d$) in terms of the corresponding  Seeley-DeWitt coefficient $B_d $\foot{ Here we ignore boundary terms. 
Note also that in the dimensional regularization 
one is to replace $\log \Lambda \to  -\frac1{{\rm d} -d}$ where $d$ is   integer  and ${\rm d} < d  $ is its analytic continuation.}
\be \la{wad}
\Gamma_{1\infty}(\Delta)= \ha  ( \log \det \Delta)_\infty  =
 - \frac{\log \Lambda}{(4\pi)^{d/2}}\,  B_d (\Delta) \ , \qquad \qquad  B_d =\int d^d  x  \, b_d (\Delta) \ . \ee
 The values   of $b_p$ for 2-derivative Laplacian $\Delta_2$ (in general curved  space and gauge field background) 
  are known up to $p=10$ (see, e.g., \cite{Gilkey:1975iq,vandeVen:1984zk,Metsaev:1987ju,Vassilevich:2003xt,vandeVen:1997pf})
 while for the 4-derivative operator  $\Delta_4$ only $b_2$ and $b_4$   were found so far 
 \ci{Gilkey:1980,Fradkin:1981iu,Barvinsky:1985an,Avramidi:2000bm}. 
 Thus to compute  the divergent part of \rf{zah} we need first to determine the 
 coefficient $b_6$ for $\Delta_4$ in \rf{zadd}. 
 This will    be the subject of the next section  and  Appendix~\ref{B}.


\section{Heat kernel coefficient \texorpdfstring{$b_6(\Delta_4)$}{b6(D4)}   }

In general, given an   elliptic   differential operator $\Delta_\rr  $ of an  even order $\rr$ in $d$ dimensions 
  one has 
\begin{equation}\label{waa}
\log \det \Delta_\rr
= - \int  \dd{^dx}  \int_\varepsilon^\infty \frac{dt}{t}\,  \tr \braket{x |  e^{-t \Delta_\rr}   |x }\ , 
\end{equation}
where tr is the trace over internal indices of the operator. 
The 
heat kernel  has  an  asymptotic expansion for $t \to  0$  so that  (see, e.g., \ci{Barvinsky:1985an,Vassilevich:2003xt,Avramidi:2000bm})
\begin{equation}\label{kac}
\tr  \braket{x | e^{-t \Delta_\rr} |x }
\equiv  \tr  K(t;x,x;\Delta_\rr) 
	\simeq \sum_{p\geq 0} \frac{2}{(4\pi)^{d/2}\,  \rr}\  t^{(p-d)/\rr} \ b_p(\Delta_\rr, d;x) \ .
\end{equation}
The  Seeley-DeWitt coefficients  $b_p$  are local invariant expressions of dimension $p$ 
constructed  out of the background    metric and gauge field, exhibiting an explicit dependence on the spacetime dimension $d$ when \(\rr \neq 2\) (below we shall consider them 
up to total derivative terms). 
In the following, we shall 
not indicate explicitly  some of the arguments of \(b_p\).
Using the proper-time cutoff $\varepsilon = \Lambda^{-\rr}$  we obtain  for the divergent part of \rf{waa}\foot{ Note that the form 
of  \eqref{kag} is universal  for any order  $\rr$  of the differential operator  -- that is the reason
 for  the above normalization of the Seeley-DeWitt coefficients.}  
\begin{equation}\label{kag}
\begin{split}
&{(\log \det \Delta_\rr)_\infty }
 =
    - \frac{2}{(4\pi)^{d/2}} \Big[
    \sum^{d-1}_{p=0} 
	\frac{{B_p(\Delta_\rr) }	}{d-p}  \Lambda^{d-p}
	+
	B_d(\Delta_\rr) \log \frac{\Lambda}{\mu} 
 \Big], 
 \\
 & B_p(\Delta_\rr)  = \int \dd{^d x} b_p(\Delta_\rr)\ .
 \end{split}
 \end{equation}
The renormalization scale $\mu$  in $\log$  will  be sometimes   left implicit below.
For example, for the  2-derivative operator defined on a vector bundle  with the  covariant derivative $\nabla_m$ and the 
curvature $\WW_{mn} = [\nabla_m, \nabla_n]$  one has\foot{ \la{nat}
Here we will  somewhat  abuse the notation and adopt  the  same labels for the connection, covariant derivative  and its curvature of the vector bundle 
as in the gauge theory $(A_m, \nabla_m, F_{mn})$  with an implicit understanding that 
  the connection in  the differential  operators $\Delta_\ell$  may take 
more general values that in a particular representation of a gauge group.}
\begin{align}
\label{kah}
& 	\Delta_2 = -\nabla^2 + X,
\end{align}
\begin{equation}
\begin{aligned}
\label{kaf}	b_6(\Delta_2) & =  	\tr\Big[
			- \frac{1}{60} \left( \covD_m \WW_{mn}  \right)^2
			+ \frac{1}{90} \WW_{mn} \WW_{nk} \WW_{km}
			- \frac{1}{12} X \WW_{mn} \WW_{mn}
			+ \frac{1}{12} X \covD^2 X
			- \frac{1}{6} X^3
		\Big].
\hspace{-1em}
\end{aligned}
\end{equation}
To find $b_6(\Delta_4)$   for the operator in \rf{zadd} 
we will  use the same idea  as in \cite{Fradkin:1981iu}   and  consider several  special cases 
of factorized   operators $ \Delta_4 $  for which 
\begin{align}\label{kae} \Delta_4 = \Delta_2\,  \Delta_2' \ , \qquad 
\det\Delta_4 = \det \Delta_2 \,  \det \Delta'_2 \ , 
\qquad 
b_d(\Delta_2 \, \Delta'_2) = b_d(\Delta_2) + b_d(\Delta'_2)  \ . 
\end{align}
This factorization Ansatz is only true for the coefficient $b_p=b_{d}$, i.e.\  with  the index $p$ equal to   the spacetime dimension $d$. 
This is related  to the fact that only the logarithmically divergent term of the expansion \eqref{kag} is universal between different regularizations, while the power-like  divergences are regularization-dependent.

The  4-derivative operator  that we   are interested in  is   given in \rf{zadd}. 
As explained in Appendix~\ref{B},   a general expression  for its  $b_6$ coefficient is  ($\hat V\equiv  \hat V_{mm} $)
\begin{equation}\label{66}
\begin{aligned}
	b_6(\Delta_4)
=
&	\tr\Big[
		\hat{k}_{ 1 }  ( \covD_m \WW_{mn}  )^2
		+ \hat{k}_{ 2 } \WW_{mn} \WW_{nk} \WW_{km} 
\\
& \hspace{2em}{}
		+ \hat{k}_{ 3 }  \hat{V}_{mn}  \hat{V}_{nk}  \hat{V}_{km}
		+ \hat{k}_{4}  \hat{V}_{mn}  \hat{V}_{mn}  \hat{V}
		+ \hat{k}_{5}  \hat{V} \hat{V} \hat{V}
		+ \hat{k}_{6}  \hat{V}_{mn} \covD_{(n} \covD_{k)}  \hat{V}_{km}
\\
& \hspace{2em}{}
		+ \hat{k}_{7}  \hat{V}_{mn} \covD^2  \hat{V}_{mn}
		+ \hat{k}_{8}   \hat{V}_{mn} \covD_{m} \covD_{n}  \hat{V}
		+ \hat{k}_{9}  \hat{V} \covD^2 \hat{V}
		+ \hat{k}_{10}  \hat{V}_{mn}  \hat{V}_{nk}  \WW_{mk}
\\
& \hspace{2em}{}
		+ \hat{k}_{11}   \WW_{mn} \covD_{(m} \covD_{k)}  \hat{V}_{kn}
		+ \hat{k}_{12}  \hat{V} \WW_{mn} \WW_{mn}
		+ \hat{k}_{13}  \hat{V}_{mn} \WW_{mk} \WW_{nk}
\\
& \hspace{2em}{}
		+ \hat{k}_{14} 	\WW_{mn} \covD_{m}  \hat{N}_n 
		+ \hat{k}_{15}  	\hat{V}_{mn} \covD_{m}  \hat{N}_{n} 
		+ \hat{k}_{16}	\hat{V} \covD_m  \hat{N}_m
		+ \hat{k}_{17}  \hat{N}_m  \hat{N}_m
		+ \hat{k}_{18}  \hat{U}\hat{V}
	\Big] \ . 
\end{aligned}
\end{equation}
As mentioned above, in contrast  to what happens in the case of $\Delta_2$ in \rf{kaf}, some of the coefficients in \rf{66}, in general,
 depend on the number of dimensions $d$. 
In the case of $d=6$ we are interested in here  one finds 
 \begin{equation}\label{67}
\begin{aligned}
\hat{k}_1   &  = - \frac{1}{30}    ,  &
\hat{k}_2   &  = \frac{1}{45}    ,  &
\hat{k}_3   &  = \frac{1}{360}    , &
\hat{k}_4   &  = \frac{1}{480}    , &
\hat{k}_5   &  = \frac{1}{2880}    , & 
\hat{k}_6   &  = - \frac{1}{120}  ,   \\
\hat{k}_7   &  = \frac{1}{120}    , &
\hat{k}_8   &  = \frac{1}{60}    , & 
\hat{k}_9   &  = \frac{1}{240}     , &
\hat{k}_{10}   &  = - \frac{1}{24}    , &
\hat{k}_{11}   &  = 0    ,  &
\hat{k}_{12}   &  = \frac{1}{24}  ,   \\
\hat{k}_{13}   &  = - \frac{1}{6}    , & 
\hat{k}_{14}   &  = - \frac{1}{3}    , & 
\hat{k}_{15}   &  = 0    , &
\hat{k}_{16}   &  = 0    , &
\hat{k}_{17}   &  = - \frac{1}{6}   ,  &
\hat{k}_{18}   &  = - \frac{1}{12}   .
\end{aligned}
\end{equation}

\section{Divergences  of 4-derivative  6d  gauge theories } 

Let us now  apply  the above general expression \rf{66}, \rf{67} for  $b_6(\Delta_4)$
to the gauge theories of interest. 

\subsection{Bosonic  theory} 
Starting with   the explicit form  of the coefficient functions   \rf{zadd}, \rf{xag}  in the operator 
$\Delta_{4A}$ and  
applying \rf{66}, \rf{67} as well as \rf{kaf},   we can compute the  coefficient $b_6$ in the 
divergent part of the effective action \rf{zah}, \rf{wad} 
of 
the  4-derivative  bosonic  6d gauge theory \rf{zaa}\foot{ In applying \eqref{66} to  the gauge field case, the trace there  is acting on the full
 internal index structure of the operator $\Delta_{4A}$, i.e.\ involving both spacetime and gauge indices (cf.\ footnote~\ref{nat}).}
\begin{align}\label{jaa}
	b_6
&=
	b_6(\Delta_{4A}) - 3  b_6 (\Delta_{2, 0})   \ ,  \ \ \ \ \ \ \ \ \ \  \\
b_6( \Delta_{4A}) & = \tr\Big[
- \frac{21}{5}  \left( \covD_m F_{mn} \right)^2
+
\Big(	\frac{2}{15} - 10 \gamma + \frac{9}{4} \gamma^3
\Big)   F_{mn} F_{nk} F_{km}\Big] \ , \label{cax}
\\
\label{cay}
b_6( \Delta_{2, 0} ) &= 	\tr\Big[ 
	- \frac{1}{60}  (\covD_m F_{mn})^2
	+ \frac{1}{90}  F_{mn} F_{nk} F_{km}
	\Big] \ . 
\end{align}
Thus finally 
\begin{align}
\label{zza}
{b}_6 &= \tr \Big[
- \frac{83}{20}   ( \covD_m F_{mn} )^2
+ \Big(
	 \frac{1}{10} - 10 \gamma + \frac{9}{4} \gamma^3
\Big)   F_{mn} F_{nk} F_{km}
\Big] \ . 
\end{align}
 Comparing to \rf{zba} we end up with the values of the one-loop $\b$-function coefficients $\b_{2A}$, $\b_{3A}$ quoted in \rf{zbb}. 
It is remarkable that the divergence proportional to $(\nabla F)^2$  turned out to be  independent of the parameter \(\gamma\):
 various terms in  $b_6$ in \eqref{66} generically do give $\gamma$-dependent   $(\nabla F)^2$
  contributions   and  they cancel out only when combined together weighted with the $\hat{k}_i$  coefficients  in \eqref{67}.

The corresponding RG equations for the renormalized  couplings $g(\mu)$ and $\g(\mu) $  
in \rf{zaa}  may be written as ($t= {1\ov (4\pi)^3}  \log \mu^2 $,  $C_2({\rm SU}(N))=N$)
\begin{align} \la{46}
&  {d g^{-2} \ov d t} =   \b_{2A} C   \ ,\qquad  \qquad {d \g  \ov d t} =     \b_{\g} C g^2 \ , \qquad \qquad 
C\equiv  \frac{1}{60} C_2 \ , \\
 \la{47}
& \b_{2A}=   249 \ , \qquad  
\b_\g   = - \gamma \beta_{2A} - \frac13 \beta_{3A} =  \frac32 (  - 2 + 34 \gamma - 45 \gamma^3)
\ . \end{align}
The flow of $g$ is independent of the parameter $\g$ and the sign of $\beta_{2A}$ corresponds to asymptotic freedom.
The fixed points of the flow of $\g$  are the   solutions of $ \b_{\g}=0$, i.e.\ 
$ \gamma_1\simeq -0.897,\ \g_2\simeq   0.059,\ \g_3\simeq  0.838$. Since 
 $  \b_\g > 0 $ for 
$  \gamma < \g_1 $  or $  \g_2  < \gamma < \g_3$,  we have that 
  $\gamma_1 $ and $\gamma_3$ are    attractive fixed points of the  flow.  
As  the sign of the $F^3$   term in \rf{zaa} is not a priori  constrained by  the requirement of positivity of the Euclidean action we 
 formally define a second coupling  $h^2=  \g^{-1}   g^{2} $ that may assume positive as well as negative values. Then   near the fixed points
$h^2$  also goes to zero in the UV, i.e.\  like $g^2$  the second coupling    is also  asymptotically free.

In Appendix \ref{C}  we shall     present also   the one-loop $\b$-functions for the  combined YM plus  4-derivative gauge  theory   with  
${\cal L}=  { 1 \ov g^2} \big[ \kappa^2 F^2 + (\nabla F)^2 +    {\g}  F^3 \big]$.

\subsection{(1,0) supersymmetric  theory}
\label{sec:hdsusy}

Let us now  consider the 6d supersymmetric version of the theory \rf{zaa}   constructed in  \cite{Ivanov:2005qf}.
In this case $\g=0$ since,  in general,  there  is no supersymmetric extension of the $F^3$  term.\foot{ This can be 
easily understood using, e.g.,  the standard $N=1$ 4d  superspace formulation: the
 YM field strength $F_{mn}$ is part of  the  spinor superfield strength $W_\alpha$  and thus 
  constructing an  invariant cubic in  $W_\alpha$   is not possible.  } 
The field content  includes 
 the 4-derivative gauge field $A_m$, the 3-derivative  6d  Weyl spinor $\Psi$,  
 and the three 2-derivative real scalars $\Phi_{I}$ ($I=1,2,3$).\foot{ In the case of the standard (1,0)  SYM  theory (corresponding to 
 $N=2$  SYM theory in 4d) the  latter  correspond to  the 
 auxiliary scalars. } In total, one has  $9+3$ bosonic and $3\times 4 $ fermionic on-shell degrees of freedom (for each  value of the internal  index).

Using  an  off-shell  harmonic superspace formulation ref.\ 
\cite{Ivanov:2005qf}  found   the   following (1,0) 
supersymmetric 6d action\footnote{ Our notation differ significantly from  that of \cite{Ivanov:2005qf}
(where, e.g.,  the scalar  kinetic term  is defined using $\epsilon^{ij}$ to raise the indices and thus implicitly  is negative definite).
  Here, the  Dirac matrices $\Gamma_m$  are $8\times8$ hermitian complex matrices satisfying $\Gamma_{(m}\Gamma_{n)}=
\ha  \{\Gamma_m, \Gamma_n  \} =  \delta_{mn}  $    and  $\Gamma_{mn} \equiv  \Gamma_{[m}\Gamma_{n]}$.
}
\begin{equation}\label{haa}
\begin{split}
S_\su  = - \frac{1}{g^2} \int  d^6 x\,  \Tr\Big[
		& \left(\covD_m {F}_{mn}\right)^2
		- i \bar \Psi \slashed{\covD} \covD^2 \Psi 
		-  \left(\nabla_m \Phi_I \right)^2
\\	&\ {}
		- \frac{i}{2}  \bar \Psi \Gamma_k \Gamma_{mn} \nabla_k\left[F_{mn} ,  \Psi \right]
		+ 2i \nabla_m F_{mn} \bar \Psi \Gamma_n \Psi
		 + \mathcal{O}\big( \Phi \Psi^2, \Phi^3\big)\Big]\ . 
\end{split}
\end{equation}
We suppressed interactions that are more than second order in the scalars and fermions, as they will  not contribute to the one-loop divergences
in a  gauge-field background. 
Note that  with our definition of the coupling constant $g$ (i.e.\ the  choice of the overall 
sign of the action)  the  gauge field term in \rf{haa} is positive definite (cf.\ \rf{zaa})  but the scalar term is not, and this is  one indication of the  non-unitarity of the theory.\footnote{ \label{fn:signg2} In \cite{Ivanov:2005qf} the opposite overall sign was  chosen  so that  their coupling is  related to ours 
by  $g^2\to-g^2$. This translates into  the  opposite sign of  the  $\b$-function  for   $g$ in \eqref{faj}. 
Note that  here there is thus no  ``preferred"  choice of the sign of the action (redefining  the scalars $\Phi_I \to i \Phi_I$
leads to  imaginary $\Phi^3$ interaction, i.e. to non-hermiticity of the action). 
For a review of related issues in higher-derivative theories see  \ci{Smilga:2017arl}.
}

The 4-derivative    operator for the fluctuations of the gauge field is given  by \rf{zadd}, \rf{xag} with $\g=0$, 
i.e.\ it is $\Delta^{(0)}_{4A} \equiv  \Delta_{4A}\big|_{\g=0}$, 
 while the 3-derivative fermion
  and the 
 2-derivative scalar
  operators  in gauge field background  may be written as\foot{  In the first form of 
$\Delta_{3\Psi}$  
the derivative in the second term acts all the way to the right 
while in the  term term it acts only on $F_{mn}$.}
\begin{equation} \la{ebe} 
\begin{split}
	& \Delta_{3 \Psi} =  i  \slashed{\covD} \covD^2 
			  + \frac{i}{2}  \slashed{\covD}
			   			\Gamma_{mn} {F}_{mn} 
			+ i \Gamma_n (\covD_m F_{mn}) = i  \slashed{\covD}^3 + i \Gamma_n (\covD_m F_{mn})\ , \qquad 
 \\&	\Delta_{2 \Phi} = - \nabla^2= \Delta_{2, 0} \ . 
\end{split}
\end{equation}
Here $ i  \slashed{\covD}^3$  is the cube of the Dirac operator \(\Delta_{ {1\Psi}}=-i\slashed{\nabla}= - i \Gamma^m \nabla_m\) whose square is
 \be  \la{412}
 \Delta_{2\Psi} = - \slashed{\nabla}^2= - \nabla^2 - \ha   \Gamma_{mn} F_{mn}  \ . \ee
As a result,  the one-loop  effective action of the supersymmetric theory \rf{haa}  is the following generalization of the bosonic case \rf{zah}
\begin{equation}\label{ccc}
\Gamma_{1\, \su} = \frac{1}{2}
		\log
		 \frac{ \det \Delta^{(0)}_{4A} \, \big[\det \Delta_{2 \Phi}\big]^3  }	 {\big[ \det \Delta_{2, 0}\big]^3\, \det \Delta_{3 \Psi} } = 
		 \frac{1}{2}
		\log \det \Delta^{(0)}_{4A}   - \ha \log  \det \Delta_{3\Psi}  \ . 
\end{equation}
Here   the contributions of the ghost and gauge-averaging  operators in \rf{zah}  got canceled against  the 
contribution of  the  three scalars $\Phi_I$. 
We also used that
 $\det \Delta_\Psi $ is  defined  for the Dirac 6d spinors   so that the factor $\ha$ accounts for the fact that the fermion  
 $\Psi$ is a Weyl  spinor.
As a result,  the coefficient of  the log divergent part  of the effective action \rf{wad}  is given by (cf.\ \rf{jaa})
\begin{align}\label{eaa}
b_{6\, \su}
	= b_6(\Delta^{(0)}_{4A}) - b_6 ( \Delta_{3\Psi} )\ .
\end{align}
Setting $\g=0$ in  \rf{cax}  gives 
\begin{equation}\label{aaz}
b_6( \Delta^{(0)}_{4A} )  = \tr\Big[
- \frac{21}{5}  \left( \covD_m F_{mn} \right)^2 + \frac{2}{15}   F_{mn} F_{nk} F_{km}\Big]\ .
\end{equation}
To compute the fermionic   contribution,  let us  first construct a 4-derivative operator 
by taking the  product of   $\Delta_{3\Psi}$ in  \rf{ebe} with  the standard Dirac operator $\Delta_{1\Psi} =-i \slashed{\nabla}$  
\begin{align}\label{fbc}
\Delta_{4\Psi}\equiv  
\Delta_{1\Psi}  \, \Delta_{3\Psi} = 
  \slashed{\nabla}^4 
+ \slashed{\covD} \Gamma_n (\covD_m F_{mn}) \ ,
& &
b_6 ( \Delta_{3\Psi} ) 
=
b_6 ( \Delta_{4\Psi} )
-
b_6 ( \Delta_{1\Psi}) \ . 
\end{align}
\(   \Delta_{4\Psi}    \) is then  a 4-order operator of the
 form \eqref{zadd} with the coefficients\footnote{ Notice that this  operator is not self-adjoint, i.e. 
 the symmetry requirements in footnote   \ref{zae}  are not satisfied.}
\begin{equation}\label{fbe}
\begin{split}
\hat{V}_{rk}  & = \Gamma_{mn} F_{mn}  \, \delta_{rk}\ ,\qquad 
\qquad
\hat{N}_k =   \frac12 \Gamma_{k} \Gamma_{n} \covD_m F_{mn}\  ,\\
\qquad 
\hat{U} & = \frac{1}{2} \Gamma_{mn}  \covD^2 F_{mn}
			 + \frac14 \Gamma_{mn} \Gamma_{rk} F_{mn} F_{rk}
			 +  \frac12 \Gamma_{k}  \Gamma_{n} \covD_k\covD_m F_{mn}\ . 
\end{split}
\end{equation}
Applying the general  expression for $b_6(\Delta_4)$  that we found in \rf{66}, \rf{67} (where now the 
 connection and its curvature   are  understood to include also  the  internal spinor  indices, see footnote~\ref{nat}) 
 and also using that squaring  $\Delta_{1\Psi}$    one obtains \eqref{412}, for which $b_6$  can  then found  from \rf{kaf},    
    we  end up with  
\begin{equation}\label{fah}
b_6( \Delta_{3\Psi} ) =  b_6 ( \Delta_{4\Psi} )
-
\ha  b_6 ( \Delta_{2\Psi})
=\tr \Big[
- \frac{8}{15}(\covD_m F_{mn})^2 
+ \frac{2}{15} F_{mn}F_{nk} F_{km}
\Big].
\end{equation}
Combining the bosonic \rf{aaz}   and the fermionic \rf{fah}  contributions to \rf{eaa} we conclude that the $F^3$ terms cancel
 as expected 
and  finally 
\begin{equation}\label{fai}
{b}_{6\, \su} = - \frac{11}{3}  \tr \left(\covD_m F_{mn}\right)^2 \, . 
\end{equation}
This is the same result as  quoted in   \rf{zba}, \rf{zbd}. 
The resulting renormalized coupling in \rf{haa}  is thus (cf.\ \rf{wad}, \rf{haa}) 
\begin{equation}\label{faj}
\frac{1}{g^2(\mu)} = \frac{1}{g^2(\Lambda)} - \frac{22}{3}  \frac{C_2 }{(4\pi)^3} \log \frac{\Lambda}{\mu} \ , 
\end{equation}
corresponding to an asymptotically free behaviour. 
This agrees with the (recently revised) result of \cite{Ivanov:2005qf} (cf.\ footnote~\ref{fn:signg2}).
Note that the computation of the $\b$-function in \cite{Ivanov:2005qf}
was done in the scalar field $\Phi_I$  background
  while here we used the gauge field background, thus providing an
independent check of the result.

For comparison, let us recall     the result  \ci{Fradkin:1982kf} 
of a  similar computation  in the ordinary-derivative (1,0)  6d SYM  theory 
\begin{equation}\label{fha}
S_{_{(1,0)\,  \SYM}}  = - \frac{\k^2}{g^2}\int d^6x \,   \Tr \Big( \ha F_{mn} F_{mn}
+ i \overline{\Psi} \slashed{\covD} \Psi -  \Phi_I \Phi_I \Big) \ , 
\end{equation}
where  $\Psi$ is a Weyl spinor, $\Phi_I$ are 3 auxiliary fields (cf. \rf{haa})
 and $\k$  is a  mass scale. 
The analog of the one-loop effective action in a gauge field background \rf{ccc} here  is 
\begin{equation}\label{418}
\Gamma_{1_{(1,0)\,  \SYM}} = \frac{1}{2}
		\log
		 \frac{ \det \Delta_{2A}  }	 {\big[ \det \Delta_{2, 0}\big]^2 
		 \, \det \Delta_{1 \Psi} } \ , \qquad \qquad 
	 (\Delta_{2A})_{mn}  =  - \delta_{mn} \covD^2 - 2 F_{mn} \ . 	 \end{equation}
Using \rf{kaf} we get 
\begin{align}
&b_6(\Delta_{2A})
		=
	\tr\Big[
		\frac{17}{30} \left( \covD_m F_{mn}  \right)^2
		+ \frac{1}{15} F_{mn} F_{nk} F_{km}
	\Big]\ , \quad 
\no
\\
& 	 b_6(\Delta_{2, 0} )
	  =
	\tr\Big[
		- \frac{1}{60} \left( \covD_m F_{mn}  \right)^2
		+ \frac{1}{90} F_{mn} F_{nc} F_{cm}
	\Big] \ ,\label{fhda}
\\
& b_6(\Delta_{1\Psi})
	 = \ha b_6(\Delta_{2\Psi}) 
	= \tr\Big[ \frac{4}{15} (\covD_m F_{mn})^2
	 + \frac{2}{45}  F_{mn} F_{nc} F_{cm}\Big] \ . 
 \no  
\end{align}
As a result, the one-loop logarithmic divergence  is  given by \rf{wad} with 
\begin{equation}\label{fhe}
b_{6_{(1,0)\,  \SYM}} =b_6 ( \Delta_{2A}) -  2  b_6 (\Delta_{2,0}) - b_6 (\Delta_{1\Psi})=
	 \frac{1}{3}\tr \left( \covD_m F_{mn}  \right)^2 \ . 
\end{equation}
Once again, the $F^3$ divergence cancels,  and  \rf{fhe} implies  the value  of $\beta_2= -20$ in \rf{zba}, \rf{15}. 
Since here 
   $ \covD_m F_{mn}=0$ is an equation of motion,  the  divergence \rf{fhe}  vanishes  on-shell, 
i.e.\ the (1,0)  6d SYM  theory is finite on-shell\foot{  The coefficient in \rf{fhe}  here is, in fact,  
gauge-dependent, see also  \ci{Buchbinder:2018lbd}.}
 though is not renormalizable off-shell.  The (1,1) 6d SYM  found by  combining the (1,0) SYM with a scalar multiplet (cf.\  \rf{15})
 is one-loop  finite even   off-shell  \ci{Fradkin:1982kf}
 (cf.\    also \ci{Bossard:2015dva}). 

Let us also note that it is  easy  to check the  cancellation of $F^3$ divergences in the (1,0) supersymmetric gauge theory \rf{haa} 
by restricting the background to satisfy $ \covD_m F_{mn}=0$ (which is  a  special on-shell   background  also in this  
theory).  Then  $\Delta_{3\Psi}$   in \rf{ebe} becomes simply  $ (\Delta_{1\Psi})^3 = i \slashed{\nabla}^3$
and also  the vector field operator in \rf{zadd}, \rf{xag} (with $\g=0$)  becomes a square 
of the standard  YM operator in \rf{418}, i.e.\ 
 $\Delta_{4A} = (\Delta_{2A})^2$.
As a result,  the effective action \rf{ccc} reduces  to 
\begin{align}\label{cccc}
\Gamma_{1\, \su}  & =  \no
	\ha 	\log \det \left(\Delta_{2A} \right)^2   -\ha   \log  \det \left(\Delta_{1\Psi} \right)^3
	\\ &	=  2\cdot \ha  \Big[ 	\log \det \Delta_{2A}  - 2\log \det \Delta_{2, 0}  - \det \Delta_{1\Psi} \Big] \no
		+ \ha   \Big[ 4  \log \det \Delta_{2, 0}  -  \det \Delta_{1\Psi} \Big] 
		\\ & = 2\, \Gamma_{1\, _{(1,0)\, \SYM}} +  \Gamma_{1\, \rm scal} \ ,  
\end{align}
i.e.\ equal to the sum of  twice the effective action of  the standard (1,0) SYM in \rf{418}    with 
the effective action of the scalar (hyper) multiplet (containing 4 real scalars and one Weyl fermion). 
Each of these do  not  contribute to  the $F^3$  divergent terms   according to \rf{15}.

\subsection*{Acknowledgments}

We  are grateful to   E. Ivanov and A. Smilga    for discussions  related to the value of the 
$\b$-function in    \cite{Ivanov:2005qf}.
  LC wishes to thank   T. Bertolini for  useful discussions. 
 AAT acknowledges K.-W. Huang and R. Roiban for  discussions   related to  \ci{Huang:2018hho}. 
LC  is supported by the International Max Planck Research School for Mathematical 
and Physical Aspects of Gravitation, Cosmology and Quantum Field Theory.  
AAT was supported by the STFC grant ST/P000762/1.

\subsection*{Note added}
After this  paper was submitted to the arXiv we learned about  the
 earlier work \ci{Gracey:2015xmw}  (see also \ci{Gracey:2016zug})
in which   a diagrammatic   computation of the two-loop $\b$-functions in the 6d  gauge theory \rf{zaa}  coupled to standard fermions 
was performed.\foot{We are grateful to  I. Klebanov  for  drawing our attention to  this paper.}
After  correcting a mistake in the   original version of this paper 
 we found that our result \rf{zbb}, \rf{zbc}  for the $\b$-functions of the theory \rf{zaa}
coupled to fermions is in full agreement 
with  the one-loop $\b$-functions  in \ci{Gracey:2015xmw}.\foot{The  translation between the  notation in  \ci{Gracey:2015xmw}
and ours is as follows.  Instead of $(\nabla_m F_{mn})^2$ in \rf{zaa} the action in    \ci{Gracey:2015xmw} contained $(\nabla_k  F_{mn})^2$ 
with the two invariants related as in footnote 3. As a result, the couplings $g_1$ and $g_2$   in \ci{Gracey:2015xmw}  are related to ours as 
$g_1 = g, \ \ g_2= 3 g(1 + \g)$ (using also that $g_2\to - g_2$  due   to  apparent  sign difference  in  notation for $F_{mn} $). 
For the  gauge theory \rf{zaa}  coupled  to Weyl fermions in generic  representation  our result  \rf{zbb}, \rf{zbc}
   for the $\b$-functions  reads (cf.\ \rf{46}, \rf{47}):
$\beta_{g} \equiv  {dg\ov dt} =  - {1\ov 120} C_2 \beta_2$, \ \ 
$\beta_{\g} \equiv  {d\g\ov dt} =  - {1\ov 120} C_2  ( 2 \g \beta_2  + {2\ov 3} \beta_3) g^2  $, \ \
$ \beta_2= 249 + N_{1\ov 2} , \    \beta_3=   9 - 900 \gamma + {405\ov 2 } \gamma^3 - 4 N_{1\ov 2}             , \ \ \ N_{1\ov 2} = {T_R\ov C_2}  N_f$.
Then  the $\beta$-functions  for the above $g_1$ and $g_2$, i.e. 
$\beta_{g_1}=  {dg_1\ov dt} = \beta_{g}, \ \   \beta_{g_2}=  {dg_2\ov dt} = 3 \beta_g (1 + \g)  +   3 g \beta_\g$
match the expressions in  \ci{Gracey:2015xmw}.
}

\iffa 
 The expressions  for the $\b$-functions disagree even in the $\g=0$ limit: 
  while  the $\gamma=0$ term  in $\beta_{2A}$ in \rf{zbb} 
or, including fermions,  $249+ 16 N_{1\ov 2}$ in $\beta_2$ in \rf{zbc}    matches the  one-loop term in  the $\beta$-function $\beta_1'\equiv \dot g_1$ in \ci{Gracey:2015xmw}, 
the corresponding  combination of  $\beta'_2=\dot g_2$  and $\beta_1'=\dot g_1$  disagrees with  the $9 - 4 N_{1\ov 2}$  term in our $\beta_3$ in \rf{zbc}.

} 
\fi

\appendix
   
\section{Derivation of the expression for   \texorpdfstring{$b_6(\Delta_4)$}{b6(D4)}   
\label{B}}

The  operator  that we shall    consider is  
\begin{align}
\label{kad}
	\Delta_4
 =
 \covD^4
			+{ V }_{mn}  \covD_m    \covD_n
			+ 2 {N}_m \covD_m  
			+ {U},
& & V_{mn} = V_{nm}\ ,	
\end{align}
which  is the most general  fourth-order elliptic differential 
operator without  3-derivative term.  It is related to the ``symmetrized'' operator in \rf{zadd}   by 
\be\la{zag}
 { V }_{mn} = \hat {V}_{mn}\ ,\quad\qquad 
	 {N}_m  =  \hat{N}_m +\ha  \covD_m \hat{V}_{mn}\ ,
	\qquad  \quad 
	U  = \tilde{U} + \covD_m \hat{N}_m \ . \ee 
The general expression  for  its  coefficient $b_6$  including only independent invariants 
   may be written as ($V\equiv  V_{mm} $)
\begin{equation}\label{64}
\begin{aligned}
	b_6(\Delta_4)
=
&	\tr\Big[
		k_{ 1 }  \left( \covD_m \WW_{mn}  \right)^2
		+ k_{ 2 } \WW_{mn} \WW_{nk} \WW_{km}
\\
& \hspace{2em}
 {}
		+ k_{ 3 }  {V}_{mn}  {V}_{nk}  {V}_{km}
		+ k_{4}  {V}_{mn}  {V}_{mn}  {V}
		+ k_{5}  {V} {V} {V}
		+ k_{6}  {V}_{mn} \covD_{(n} \covD_{k)}  {V}_{km}
\\
& \hspace{2em}{}
		+ k_{7}  {V}_{mn} \covD^2  {V}_{mn}
		+ k_{8}   {V}_{mn} \covD_{m} \covD_{n}  {V}
		+ k_{9}  {V} \covD^2 V
		+ k_{10}  {V}_{mn}  {V}_{nk}  \WW_{mk}
\\
& \hspace{2em}
 {}
		+ k_{11}   \WW_{mn} \covD_{(m} \covD_{k)}  {V}_{kn}
		+ k_{12}  {V} \WW_{mn} \WW_{mn}
		+ k_{13}  {V}_{mn} \WW_{mk} \WW_{nk}
\\
& \hspace{2em}
		{}+ k_{14} 	\WW_{mn} \covD_{m}  {N}_n 
		+ k_{15}  	V_{mn} \covD_{m}  {N}_{n} 
		+ k_{16}	V \covD_m  {N}_m
		+ k_{17}  {N}_m  {N}_m
		+ k_{18}  {UV}
	\Big] , 
\end{aligned}
\end{equation}
where 
the trace is over internal indices and
 $k_i$ 
 are real coefficients.\footnote{
The relations between the $k_i$  and $\hat k_i$ in \rf{67}   are, using \eqref{zag},
$
\hat{k}_6  =  k_6 + \frac12 k_{15} - \frac14 k_{17}, \
\hat{k}_8  =  k_8 + \frac12 k_{16}, \
\hat{k}_{10}  =  k_{10} - \frac12 k_{15} + \frac14 k_{17}, \
\hat{k}_{11} =  k_{11} + \frac12 k_{14}, \
\hat{k}_{15} =  k_{15} - k_{17}, \
\hat{k}_{16} =  k_{16}  + k_{18}
$
with   $\hat k_i= k_i$ otherwise.} 
Their values  in $d=6$  found below are 
 \begin{equation}\label{65}
\begin{aligned}
k_1   &  = - \frac{1}{30}  ,  &  
k_2   &  = \frac{1}{45}   , &  
k_3   &  = \frac{1}{360}  ,  & 
k_4   &  = \frac{1}{480}  ,  & 
k_5   &  = \frac{1}{2880}  ,  & 
k_6   &  = \frac{1}{30}    , \\
k_7   &  = \frac{1}{120}  ,  & 
k_8   &  = - \frac{1}{40}  ,  &  
k_9   &  = \frac{1}{240}   ,  & 
k_{10}   &  = - \frac{1}{12},    & 
k_{11}   &  = \frac{1}{6}   , & 
k_{12}   &  = \frac{1}{24}   ,  \\
k_{13}   &  = - \frac{1}{6}  ,  &  
k_{14}   &  = - \frac{1}{3}  ,  &  
k_{15}   &  = - \frac{1}{6}  ,  & 
k_{16}   &  = \frac{1}{12}  ,  & 
k_{17}   &  = - \frac{1}{6}  , &  
k_{18}   &  = - \frac{1}{12}  . 
\end{aligned}
\end{equation}
To determine $k_i$ we shall exploit the factorization property  \rf{kae},  i.e. 
\be b_6(\Delta_4) =  b_6 (\Delta_2) + b_6 (\Delta'_2) \ , \ \ \ \qquad \ \ \   \Delta_4 = \Delta_2 \Delta_2' \ , \la{rabb}  \ee 
where $b_6(\Delta_2)$   is given by  \eqref{kaf}.
As was already remarked in Section~3, such factorization applies here because we are considering  $b_6$ in 6 spacetime dimensions. 

One needs  to identify enough special cases and consistency conditions 
to fix all $k_i$. 
When comparing the two sides of the $b_6$-relation in  \rf{rabb}    it is important to take into account 
  (i) that they are defined up to total derivatives (which we drop in discussing UV divergences), 
 (ii) that the terms can be cyclically permuted because they appear  under  an overall trace, 
 and  (iii)  relations between  the invariants (implied, e.g.,   by the   Bianchi identity).

Considering \( \Delta_2 = - \nabla^2 + X \) and \( \Delta_2'=  - \nabla^2 + X' \)  their product is given by \rf{kad}  with 
\begin{align}\label{rac}
V_{mn} = - \delta_{mn} (X+X') ,
\quad 
N_m = - \covD_m X' ,\quad 
U = XX' - \covD^2 X' , \quad   V = - 6 (X+X') \ . 
\end{align}
Using \rf{kaf}   and comparing with \rf{64}    gives 
\begin{equation}\label{rad}
\begin{aligned}
 k_1  & = - \frac{1}{30},  &
  k_3 + 6 k_4 + 36 k_5  & = \frac{1}{36},  &
 \ \ \ \  k_{13} + 6 k_{12} & = \frac{1}{12},   &  
\ \ \   k_{17}  & = - \frac{1}{6},
\\
 k_2  & = \frac{1}{45},  &
  k_6 + 6 k_7 + 6 k_8 + 36 k_9 &  = \frac{1}{12 }, & 
\ \ \ \  k_{15} + 6 k_{16} & = \frac{1}{3},   &  
   \ \ \  k_{18}  & = - \frac{1}{12}.
\end{aligned}
\end{equation}
Next, let us assume that  
\begin{align}\label{rae}
\Delta_4 & = \Delta_+ \Delta_- , \qquad \nabla_m^\pm \equiv \nabla_m \pm  K_m \ ,
\\
 \Delta_\pm  & = -(\nabla_m^\pm)^2 =- \nabla^2 \mp 2 K_m \covD_m  \mp (\covD_m K_m) - K_m K_m 
 \ . 
\end{align}
Here   $\nabla_m K_n= \partial_m K_n  + [A_m, K_n]  $  ($K_m$ is  in the adjoint representation of the gauge group). 
The coefficient functions  in  the  corresponding operator $\Delta_4=  \Delta_+ \Delta_- $  in \eqref{kad} read
\begin{equation}\label{rah}
\begin{split}
V_{mn} & = - 4 \covD_{(m} K_{n)} + 2 K^2 \delta_{mn} - 4 K_{(m} K_{n)}, \qquad \qquad V = -4 \covD_n K_n  +8 K^2\ ,  \\
N_m & = - \covD^2 K_m - \covD_m \covD_n K_n + \covD_m K^2 
\\
&\hspace{3em} + K_m K^2 - K^2 K_m - 2 K_n \covD_n K_m - K_m \covD_n K_n + 2 K_n \WW_{nm} , \\
U & = - \covD^2 \covD_n K_n + \covD^2 K^2 - 2 K_m \covD_m \covD_n K_n + 2 K_m \covD_m K^2 - (\covD_n K_n)^2 + K^4
\\
&\hspace{3em}   + (\covD_n K_n)K^2 - K^2 \covD_n K_n -2 \covD_m K_n \WW_{mn} - 2 K_m K_n \WW_{mn}  + 2 K_m \covD_n \WW_{mn}.
\end{split}
\end{equation}
Using \rf{kaf}  and the relations 
\begin{equation}\label{raf}
\WW^{\pm}_{mn} \equiv  [\covD^\pm_m,\covD^\pm_n] = \WW_{mn} + [K_m, K_n] \pm (\covD_m K_n - \covD_n K_m)\ , 
\end{equation}
\begin{equation}\label{rag}
\begin{aligned}
	\covD^\pm_m \WW_{mn}^\pm
= 
	\covD_m & \Big[ \WW_{mn} + [K_m, K_n] \pm (\covD_m K_n - \covD_n K_m)^{\vphantom A} \Big]\\
&	\pm \Big[ K_m, \WW_{mn} + [K_m, K_n] \pm (\covD_m K_n - \covD_n K_m)^{\vphantom A}  \Big] \ , 
\end{aligned}
\end{equation}
 one can compute $b_6(\Delta_\pm)$ and then  compare to $b_6(\Delta_4)$ in \rf{rabb}. 
 
It is  enough to consider the following  special cases: 
\begin{enumerate}
\item Abelian gauge group,  $\nabla_n  K_m =   \partial_n K_m$,   $[\WW_{mn}, K_k]=0$.
 In  \eqref{rabb} we consider the terms  with  $\partial^r K_m$, $r=0, 1,4$
that can always be uniquely cast into the form
\begin{align}
K^6,
& &
K^4 \partial_m K_m \ ,
& &
(\partial_m K_m )\partial^2(\partial_n K_n )\ ,
& &
K_m \partial^4 K_m \ .
\end{align}
Then  comparing also the coefficients of \( \WW_{nm} K^2 \partial_n K_m \) and \( \WW_{nm} K_m \partial_n K^2 \) (the latter does not actually appear)  one obtains  
 \begin{equation}\label{rai}
\begin{aligned}
&32 k_3 + 16\cdot 12 k_4 + 512 k_5 + 8 k_{18}  = 0\ , \qquad 
   k_{15} + 4 k_{16} - k_{17} + 4 k_{18}  = 0 \ ,   \\
&4 k_6 + 8 k_7 + 2 k_{15} - k_{17}   =  \frac{1}{30}\ , \qquad
12 k_{18} + 256\cdot 3 k_{5} + 16 \cdot 18 k_4 + 48 k_3  = 0 \ , \qquad  \\
&12 k_6 + 8 k_7 + 16k_8 + 16 k_9 + 6 k_{15} + 8 k_{16} - 3 k_{17} + 4 k_{18}  = - \frac{1}{30}  \ . 
\end{aligned}
\end{equation}
\item  $K_n$  constrained by  $\nabla_m K_n =0, $ implying 
\( 
	2 \covD_{[k} \covD_{m]} K_n 
= 
	[\WW_{km} , K_n ]=0 .
\)
 This leads to  a number of nontrivial relations, e.g., \ \( \tr ([K_m, K_n] \WW_{nk} \WW_{km} )= 0 \).
 All the remaining invariants  can be uniquely written as a combination of
\begin{equation}
\begin{aligned}
& K^6\ , \qquad  K_m K_n K_k K_m K_n K_k \ , \qquad   K^2 K_m K^2 K_m\ ,  \qquad K_m K_n K_m K_k K_n K_k \ , \\
& K^2 \WW_{mn} \WW_{mn}\ ,  \qquad   K^2 K_m K_n K_m K_n \ ,   \qquad   K_m K_n \WW_{mk} \WW_{nk}\ , \qquad F_{mn} K_m K_n K^2 \ . 
\end{aligned}
\end{equation}
Their coefficients can then be compared to get ($(K_m K_n K_p)^2$ and $(K_m K_p K_m)^2$ give the same equation)
 \begin{equation}\label{raj}
\begin{aligned}
 24 k_{3}  = & \frac{1}{15} \ ,  \qquad 
-8 k_3 + 64 k_4 + 512 k_5 + 8 k_{18} - 2 k_{17}  = - \frac{2}{45} \ ,    \\
 k_{13} - k_{17}  = 0 \ , & \qquad 
64 k_{4} + 48 k_{3} + 2 k_{17} = -  \frac{1}{15}\ , \qquad 
24 k_3 + 64 k_4   =  \frac{1}{5} \ ,\\
& - k_{17} +2 k_{18}  = 0 \ ,  \qquad  
8 k_{12} + 2 k_{13}  =  0 \ .
\end{aligned}
\end{equation}
\item 
General  unconstrained  $K_n$, comparing   the terms with one $K_m$ or two of them contracted together.
A basis of such tensors  contains 
\begin{equation}\label{rak}
\begin{aligned}
& K_m \covD^2 \covD_n \WW_{mn} ,\qquad
& K_m F_{nk} \nabla_n F_{km} ,\qquad
& K_m F_{mn} \nabla_k \WW_{kn}  , \\
& K_m \covD_n \WW_{km}  \WW_{nk} ,\qquad
& K_m   \covD_k F_{kn} \WW_{mn} ,\qquad
& K_m \covD^4 K_m ,\\
& K_m  \covD_k \WW_{kn}  \covD_n K_m,\qquad
& K^2  \WW_{kn} \WW_{kn} ,\qquad
& K_m  \WW_{kn} K_m \WW_{kn}.
\end{aligned}
\end{equation}
In this case we obtain (the two $KKFF$ terms  give the same equation)
 \begin{equation}\label{ral}
\begin{aligned}
&2 k_{11}+ k_{14}  =  0 \ , \qquad 
8 k_{12} + 2 k_{13}  = 0  \ , \qquad 
2 k_{11} - 2 k_{13} + 2 k_{14}   = 0 \ ,  \\
&4 k_6 + 8 k_7 +2  k_{15} - k_{17}  =  \frac{1}{30}   \ , \qquad 
2k_6 + 16 k_7 + 2 k_{10} + 8k_{12} + 2 k_{13}  =  \frac{1}{30}\ ,   \\
&4k_6 + 16 k_7 + 4 k_{10}  = -\frac{1}{15}\ ,   \qquad   
2k_6 + 16 k_7 + 2 k_{10}  =  \frac{1}{30}  \  , \qquad 
k_{11} + k_{13}  = 0\ . 
\end{aligned}
\end{equation}
\end{enumerate}
The final  system of equations  is given  by  \eqref{rad}, \eqref{rai}, \eqref{raj} and \eqref{ral}. 
 This  system is over-determined, with the unique solution  for $k_i$  given  by \eqref{65}. 
 That some of the equations  are actually redundant   gives a non-trivial consistency check of the calculation.
  We also checked some of the  coefficients $k_i$ 
 by explicit  diagrammatic calculations of the corresponding UV divergences. 

\section{One-loop divergences in  $F^2 + (\nabla F)^2 + F^3 $ theory}
\la{C}
It is   straightforward to generalize the expression  for the effective action \rf{zah} 
to the case when  one  adds to the action \rf{zaa}   the standard YM term, i.e. the first term in \rf{fha}
\be \la{b1}
 \Gamma_{1}
  =
\frac{1}{2} \log \frac{\det \Delta'_{4A} }{\big[\det\left(-\nabla^2\right)\big]^2 \det\left(-\nabla^2 + \k^2\right) }   
\ ,   \qquad \qquad  \Delta'_{4A} = \Delta_{4A} + \k^2 \Delta_{2A} \ .   \ee
Here $\Delta_{2A}$ is given in \rf{418}. 
The  quadratic  and logarithmic  divergences of \rf{b1} 
 are determined  by the total $b_4$ and $b_6$ coefficients (cf.\ \rf{kag})
\begin{align}\label{raa}
  & \Gamma_{1\infty } = 
    - \frac{1}{(4\pi)^{3}} \Big(
\ha B_4  \Lambda^{2}
	+
	B_6 \log\Lambda 	\Big) \ ,\\
	&   B_p = \int\dd{^6x} b_p \, , \qquad \qquad 
b_p =  b_p(  \Delta'_{4A}   ) - 2\, b_p (- \nabla^2) -  b_p (- \nabla^2+\k^2) \ . \la{b3}
\end{align}
The expression for   $b_4$ is known for both for $\Delta_2$  \rf{kah}  and $\Delta_4$ \rf{kad} 
operators \cite{Gilkey:1980, Gusynin:1988zt}\footnote{
Here tr  and $F_{mn}$ are the general trace and the curvature on the bundle, cf.\ footnote \ref{nat}.
The factor \(\sqrt \pi \)  comes from an overall  \( \Gamma[\frac d4]\) in the  expression of the heat kernel coefficient $b_4(\Delta_4,d)$.  }
\begin{align}\label{rabv}
& b_4(\Delta_2)  = \tr \Big[\frac{1}{12} F_{mn} F_{mn} + \frac12 X^2 \Big]\ , 
\\
& b_4(\Delta_4, d=6)  = \sqrt{\pi} \tr \Big[\frac{1}{12} F_{mn} F_{mn} +  \frac{1}{16}
 V_{mn} V_{mn} 
 + \frac{1}{32}VV - \frac 14 U \Big]\ . \la{77b}   
\end{align}
The coefficient 
$b_4$  controls the logarithmic divergences in the corresponding 4d  theory 
 where their computation was done in \ci{Fradkin:1981iu} (see also \ci{Casarin:2017xez}). 
For the operators in \rf{b1}  we get in $d=6$ 
 (here $\tr$  is in the adjoint representation and $F_{mn}$ is the gauge field strength)\footnote{
 $b_4(\Delta'_{4A}) $
 has two sources of dependence on the space-time dimension $d$:   the operator itself   and  the  coefficients
  in $b_4$  in \rf{77b}.
 The gauge fixing   contributions are independent of $d$.
As a result, 
in 4d theory the coefficient  $\beta_{1A}$ in \rf{b6}  below  is given by  
$\beta_{1A}  = -2 ( 43  + 108 \gamma + 27 \gamma^2 )$ (cf. \ci{Fradkin:1981iu,Casarin:2017xez}). }
\begin{align}\label{raca}
 &b_4(- \nabla^2 +\k^2)  = \frac{1}{12} \tr F_{mn} F_{mn} + \frac12\, \k^4\, C_2\ , \\
&
 b_4(\Delta'_{4A})  =  -   \Big(\frac 32 +\frac 92 \gamma + \frac 9 8 \gamma^2  \Big) \sqrt{\pi} \tr F_{mn} F_{mn} + \frac 9 8 \sqrt \pi\, \k^4\, C_2 \ . \la{rcca} 
\end{align}
Similarly, using \rf{kaf}   and \rf{66}, \rf{67} we find 
\begin{align}
\nonumber
& b_6( \Delta'_{4A} )  = 
 - \frac{21}{5} \tr  (\covD_m F_{mn})^2 
+ \Big(
 \frac{2}{15} - 10 \gamma + \frac{9}{4} \gamma^3
\Big)  \tr F_{mn} F_{nk} F_{km}
\\ 
\label{caxy} & \hspace{6em} {}
+ \Big(
	\frac32 + 9 \g + 3 \g^2
\Big) \k^2 \tr F_{mn} F_{mn}
-   \k^6 C_2 \ , 
\\
\label{cayf}
& b_6( -\nabla^2 + \k^2  ) = 	
	- \frac{1}{60} \tr  (\covD_m F_{mn})^2
	+ \frac{1}{90}  \tr F_{mn} F_{nk} F_{km}
	- \frac{1}{12} \k^2  \tr F_{mn} F_{mn}
	- \frac{9}{8}\k^6 C_2  \ . 
\end{align}
As a result, the total values  of the  coefficients of the quadratic and logarithmic divergences in \rf{raa}  in $d=6$   are 
(omitting field-independent terms)
\begin{align}
\la{bb4} 
&  b_4 =  \frac{1}{12} \beta_{1}   \tr F_{mn} F_{mn} \ , 
 \\
 & b_6 = \k^2 \beta_{\k} \tr F_{mn} F_{mn}
- \frac{1}{60}\beta_{2A}
 \tr  (\covD_m F_{mn})^2 
+ \frac{1}{90}\beta_{3A}  \tr F_{mn} F_{nk} F_{km} \ ,   \la{66b}
\\ 
& \beta_{1A}  =-  3 -18\sqrt \pi - 54\sqrt\pi \gamma - {27\ov 2} \sqrt \pi \gamma^2  \ ,
 \qquad  \qquad 
 \beta_{\k,A} =
			\frac{19}{12} + 9\g + 3 \g^2   \  , 
		 \la{b6}
\end{align} 
where $\beta_{2A}$ and $\beta_{3A}$  in \rf{66b} are the same as in \rf{zbb}. 
Ignoring   non-universal  quadratic divergence (absent in dimensional regularization), the  logarithmic renormalization of $\kappa$  is controlled by $\beta_{\k,A}$
with the RG equation (cf.\ \rf{46}, \rf{47})\foot{ Recall that the coefficient of the YM  term is chosen as $\k^2\ov g^2$, cf.\ \rf{fha}. } 
\begin{equation}\label{b12}
{ d \k^2 \ov d t} = -  \Big(\b_{\k,A} + \frac{1}{60}\beta_{2A}\Big)2\k^2 g^2 {C_2} 
 = \Big( - \frac{172}{15} - 18 \gamma  -6 \gamma ^2\Big) \k^2 g^2 {C_2}
 \ . 
\end{equation}
Near both the attractive fixed points $ \gamma_1 \simeq - 0.897$ and  $ \gamma_3 \simeq 0.838$    of $\beta_\g$ in \rf{47}, the r.h.s\  of \rf{b12} is negative and thus
$\k^2 \to 0$  in the UV.

Let us now consider the log divergence in  the (1,0)  supersymmetric extension of this  bosonic model, i.e.\ the 
$(1,0)$ SYM  combined with the (1,0) theory \rf{haa}. Here the operators in the 1-loop  effective action \rf{ccc} 
get $\k$-dependent terms  as  in  \rf{b1} (with  $\g=0$)
\be\la{b13}
\Delta'^{(0)}_{4A} = \Delta^{(0)}_{4A} + \kappa^2 \Delta_{2A}\ , 
\qquad 
\Delta_{3\Psi}' = \Delta_{3\Psi} + \kappa^2 \Delta_{1\Psi}\ , 
		\qquad 
\Delta_{2\Phi}' = \Delta_{2\Phi} + \kappa^2\ , 
\ee 
where $\Delta_{1\Psi} = - i \slashed{\nabla}$  and $\Delta_{2\Phi} = - \nabla^2$.
 Explicitly,  we get (cf.\ \rf{ccc}, \rf{b1})
\be
\begin{aligned}
\label{qad}
\Gamma'_{1\su} & =  \frac{1}{2}
	\log \Big[
			 \frac{\det \Delta'^{(0)}_{4A}}{[\det (-\nabla^2)]^2 \det (-\nabla^2 + \kappa^2)}
			 \frac{[\det \Delta'_{2\Phi}]^3 }{\det \Delta'_{3\Psi}}
		\Big]
		\\
& 	=  \frac{1}{2}
			\log \Big[
					 \frac{\det \Delta'^{(0)}_{4A}}{\det \Delta'_{3\Psi}}
					 \frac{[\det (-\nabla^2 + \kappa^2)]^2 }{[\det (-\nabla^2)]^2 }
				\Big]\ . 
\end{aligned}
\ee
For the gauge field and  scalar determinants   the  expressions for $b_4$ and $b_6$  are 
 given by  \eqref{raca}--\eqref{cayf}  with $\g=0$   while for 
 the fermion   contribution we get as in \eqref{fbc},
\begin{equation}\label{qag}
b_6 (\Delta_{3\Psi}') = b_6( \Delta_{1\Psi}\Delta_{3\Psi}') -  b_6( \Delta_{1\Psi})=
 b_6(\Delta_{3\Psi}) +  \frac{14}{3}\kappa^2 \tr F_{mn} F_{mn} - \frac43 \kappa^6 C_2.
\end{equation}
As a result, the analog of \rf{b6} is 
\be
\label{qah}
\begin{aligned}
& b_6 = \kappa^2 \beta_{\kappa\,  (1,0)} \tr F_{mn} F_{mn} - \frac{1}{60} \beta_{2\,(1,0)}  \tr \left(\nabla_m F_{mn}\right)^2
\ ,  
\\
& \beta_{\kappa \,(1,0)} = - \frac{29}{6} \, ,\qquad \qquad 
\beta_{2\, (1,0)} = 220\, .
\end{aligned}
\ee
where $\beta_{2\, (1,0)}$  is the same as in \rf{fai}, \rf{zbd}. 
Since the combination $\beta_{\kappa \,(1,0)} + \frac{1}{60} \beta_{2A} $ is negative, as a result of \eqref{b12} we do not have asymptotic freedom in the supersymmetric case. 

As a final comment,  notice also that on $\nabla_m F_{mn}=0$ background \rf{qad}  becomes the following generalization   of \rf{cccc}
\begin{align}\label{ccc1}
\Gamma'_{1\, \su}   	= &\ \ \  \ha  \Big[ 	\log \det \Delta_{2A}  - 2\log \det \Delta_{2, 0}  - \det \Delta_{1\Psi} \Big]  \no \\ 
&+  \ha  \Big[ 	\log \det( \Delta_{2A} + \k^2 )  - 2\log \det (\Delta_{2, 0} + \k^2)  - \det (\Delta_{1\Psi}  + \k) \Big] \no \\ 
&+ \ha   \Big[ 4  \log \det ( \Delta_{2, 0} + \k^2)  -  \det (\Delta_{1\Psi} + \k)\Big] 
\  ,   
\end{align}
i.e. the sum of  contributions of  massless  (1,0) SYM, its massive analog, and a  massive  analog of scalar  multiplet. 


\bibliography{biblio} 

\providecommand{\href}[2]{#2}\begingroup\raggedright\begin{thebibliography}{10}

\bibitem{Fradkin:1982kf}
E.~S. Fradkin and A.~A. Tseytlin, \emph{{Quantum properties of higher
  dimensional and dimensionally reduced supersymmetric theories}},
  \href{http://dx.doi.org/10.1016/0550-3213(83)90022-6}{\emph{Nucl. Phys.} {\bf
  B227} (1983) 252}.

\bibitem{Fradkin:1981iu}
E.~S. Fradkin and A.~A. Tseytlin, \emph{{Renormalizable asymptotically free
  quantum theory of gravity}},
  \href{http://dx.doi.org/10.1016/0550-3213(82)90444-8}{\emph{Nucl. Phys.} {\bf
  B201} (1982) 469--491}.

\bibitem{Grinstein:2008qq}
B.~Grinstein and D.~O'Connell, \emph{{One-Loop Renormalization of Lee-Wick
  Gauge Theory}},
  \href{http://dx.doi.org/10.1103/PhysRevD.78.105005}{\emph{Phys. Rev.} {\bf
  D78} (2008) 105005}, [\href{http://arxiv.org/abs/0801.4034}{{\tt
  0801.4034}}].

\bibitem{Casarin:2017xez}
L.~Casarin, \emph{{On higher-derivative gauge theories}},  2017,
  [\href{https://arxiv.org/abs/1710.08021}{\tt 1710.08021}].

\bibitem{Ivanov:2005qf}
E.~A. Ivanov, A.~V. Smilga and B.~M. Zupnik, \emph{{Renormalizable
  supersymmetric gauge theory in six dimensions}},
  \href{http://dx.doi.org/10.1016/j.nuclphysb.2005.08.014}{\emph{Nucl. Phys.}
  {\bf B726} (2005) 131--148}, [\href{http://arxiv.org/abs/hep-th/0505082}{{\tt
  hep-th/0505082}}].

\bibitem{Smilga:2006ax}
A.~V. Smilga, \emph{{Chiral anomalies in higher-derivative supersymmetric 6D
  theories}},
  \href{http://dx.doi.org/10.1016/j.physletb.2007.02.002}{\emph{Phys. Lett.}
  {\bf B647} (2007) 298--304}, [\href{http://arxiv.org/abs/hep-th/0606139}{{\tt
  hep-th/0606139}}].

\bibitem{Beccaria:2015uta}
M.~Beccaria and A.~A. Tseytlin, \emph{{Conformal a-anomaly of some non-unitary
  6d superconformal theories}},
  \href{http://dx.doi.org/10.1007/JHEP09(2015)017}{\emph{JHEP} {\bf 09} (2015)
  017}, [\href{http://arxiv.org/abs/1506.08727}{{\tt 1506.08727}}].

\bibitem{Beccaria:2015ypa}
M.~Beccaria and A.~A. Tseytlin, \emph{{Conformal anomaly c-coefficients of
  superconformal 6d theories}},
  \href{http://dx.doi.org/10.1007/JHEP01(2016)001}{\emph{JHEP} {\bf 01} (2016)
  001}, [\href{http://arxiv.org/abs/1510.02685}{{\tt 1510.02685}}].

\bibitem{Giombi:2015haa}
S.~Giombi, I.~R. Klebanov and G.~Tarnopolsky, \emph{{Conformal QED$_d$,
  $F$-Theorem and the $\epsilon$ Expansion}},
  \href{http://dx.doi.org/10.1088/1751-8113/49/13/135403}{\emph{J. Phys.} {\bf
  A49} (2016) 135403}, [\href{http://arxiv.org/abs/1508.06354}{{\tt
  1508.06354}}].

\bibitem{Giombi:2016fct}
S.~Giombi, G.~Tarnopolsky and I.~R. Klebanov, \emph{{On $C_{J}$ and $C_{T}$ in
  Conformal QED}}, \href{http://dx.doi.org/10.1007/JHEP08(2016)156}{\emph{JHEP}
  {\bf 08} (2016) 156}, [\href{http://arxiv.org/abs/1602.01076}{{\tt
  1602.01076}}].

\bibitem{Osborn:2016bev}
H.~Osborn and A.~Stergiou, \emph{{C$_{T}$ for non-unitary CFTs in higher
  dimensions}}, \href{http://dx.doi.org/10.1007/JHEP06(2016)079}{\emph{JHEP}
  {\bf 06} (2016) 079}, [\href{http://arxiv.org/abs/1603.07307}{{\tt
  1603.07307}}].

\bibitem{Huang:2018hho}
K.-W. Huang, R.~Roiban and A.~A. Tseytlin, \emph{{Self-dual 6d 2-form fields
  coupled to non-abelian gauge field: quantum corrections}},
  \href{http://dx.doi.org/10.1007/JHEP06(2018)134}{\emph{JHEP} {\bf 06} (2018)
  134}, [\href{http://arxiv.org/abs/1804.05059}{{\tt 1804.05059}}].

\bibitem{Johansson:2018ues}
H.~Johansson, G.~Mogull and F.~Teng, \emph{{Unraveling conformal gravity
  amplitudes}}, \href{http://dx.doi.org/10.1007/JHEP09(2018)080}{\emph{JHEP}
  {\bf 09} (2018) 080}, [\href{http://arxiv.org/abs/1806.05124}{{\tt
  1806.05124}}].

\bibitem{Ivanov:2005kz}
E.~A. Ivanov and A.~V. Smilga, \emph{{Conformal properties of hypermultiplet
  actions in six dimensions}},
  \href{http://dx.doi.org/10.1016/j.physletb.2006.05.003}{\emph{Phys. Lett.}
  {\bf B637} (2006) 374--381}, [\href{http://arxiv.org/abs/hep-th/0510273}{{\tt
  hep-th/0510273}}].

\bibitem{Kuzenko:2017jdy}
S.~M. Kuzenko, J.~Novak and S.~Theisen, \emph{{New superconformal multiplets
  and higher derivative invariants in six dimensions}},
  \href{http://dx.doi.org/10.1016/j.nuclphysb.2017.10.013}{\emph{Nucl. Phys.}
  {\bf B925} (2017) 348--361}, [\href{http://arxiv.org/abs/1707.04445}{{\tt
  1707.04445}}].

\bibitem{Kuzenko:2017xgh}
S.~M. Kuzenko, J.~Novak and I.~B. Samsonov, \emph{{Chiral anomalies in six
  dimensions from harmonic superspace}},
  \href{http://dx.doi.org/10.1007/JHEP11(2017)145}{\emph{JHEP} {\bf 11} (2017)
  145}, [\href{http://arxiv.org/abs/1708.08238}{{\tt 1708.08238}}].

\bibitem{Gilkey:1975iq}
P.~B. Gilkey, \emph{{The Spectral geometry of a Riemannian manifold}},
  \href{http://dx.doi.org/10.4310/jdg/1214433164}{\emph{J. Diff. Geom.} {\bf
  10} (1975) 601--618}.

\bibitem{Fradkin:1983tg}
E.~S. Fradkin and A.~A. Tseytlin, \emph{{Conformal Anomaly in Weyl Theory and
  Anomaly Free Superconformal Theories}},
  \href{http://dx.doi.org/10.1016/0370-2693(84)90668-3}{\emph{Phys. Lett.} {\bf
  134B} (1984) 187}.

\bibitem{Fradkin:1985am}
E.~S. Fradkin and A.~A. Tseytlin, \emph{{Conformal Supergravity}},
  \href{http://dx.doi.org/10.1016/0370-1573(85)90138-3}{\emph{Phys. Rept.} {\bf
  119} (1985) 233--362}.

\bibitem{Jack:1982hf}
I.~Jack and H.~Osborn, \emph{{Two Loop Background Field Calculations for
  Arbitrary Background Fields}},
  \href{http://dx.doi.org/10.1016/0550-3213(82)90212-7}{\emph{Nucl. Phys.} {\bf
  B207} (1982) 474--504}.

\bibitem{Jack:1983sk}
I.~Jack and H.~Osborn, \emph{{Background Field Calculations in Curved
  Space-time. 1. General Formalism and Application to Scalar Fields}},
  \href{http://dx.doi.org/10.1016/0550-3213(84)90067-1}{\emph{Nucl. Phys.} {\bf
  B234} (1984) 331--364}.

\bibitem{Grinstein:2015ina}
B.~Grinstein, A.~Stergiou, D.~Stone and M.~Zhong, \emph{{Two-loop
  renormalization of multiflavor $\phi^3$ theory in six dimensions and the
  trace anomaly}},
  \href{http://dx.doi.org/10.1103/PhysRevD.92.045013}{\emph{Phys. Rev.} {\bf
  D92} (2015) 045013}, [\href{http://arxiv.org/abs/1504.05959}{{\tt
  1504.05959}}].

\bibitem{vandeVen:1984zk}
A.~E.~M. van~de Ven, \emph{{Explicit Counter Action Algorithms in Higher
  Dimensions}},
  \href{http://dx.doi.org/10.1016/0550-3213(85)90496-1}{\emph{Nucl. Phys.} {\bf
  B250} (1985) 593--617}.

\bibitem{Metsaev:1987ju}
R.~R. Metsaev and A.~A. Tseytlin, \emph{{On loop corrections to string theory
  effective actions}},
  \href{http://dx.doi.org/10.1016/0550-3213(88)90306-9}{\emph{Nucl. Phys.} {\bf
  B298} (1988) 109--132}.

\bibitem{Vassilevich:2003xt}
D.~V. Vassilevich, \emph{{Heat kernel expansion: User's manual}},
  \href{http://dx.doi.org/10.1016/j.physrep.2003.09.002}{\emph{Phys. Rept.}
  {\bf 388} (2003) 279--360}, [\href{http://arxiv.org/abs/hep-th/0306138}{{\tt
  hep-th/0306138}}].

\bibitem{vandeVen:1997pf}
A.~E.~M. van~de Ven, \emph{{Index free heat kernel coefficients}},
  \href{http://dx.doi.org/10.1088/0264-9381/15/8/014}{\emph{Class. Quant.
  Grav.} {\bf 15} (1998) 2311--2344},
  [\href{http://arxiv.org/abs/hep-th/9708152}{{\tt hep-th/9708152}}].

\bibitem{Gilkey:1980}
P.~B. Gilkey, \emph{{The spectral geometry of the higher order Laplacian}},
  \href{http://dx.doi.org/10.1215/S0012-7094-80-04731-6}{\emph{Duke Math. J.}
  {\bf 47} (1980) 511--528 [Err: 48 (1981) 887]}.

\bibitem{Barvinsky:1985an}
A.~O. Barvinsky and G.~A. Vilkovisky, \emph{{The Generalized Schwinger-Dewitt
  Technique in Gauge Theories and Quantum Gravity}},
  \href{http://dx.doi.org/10.1016/0370-1573(85)90148-6}{\emph{Phys. Rept.} {\bf
  119} (1985) 1--74}.

\bibitem{Avramidi:2000bm}
I.~G. Avramidi, \emph{{Heat kernel and quantum gravity}},
  \href{http://dx.doi.org/10.1007/3-540-46523-5}{\emph{Lect. Notes Phys.
  Monogr.} {\bf 64} (2000) 1--149}.

\bibitem{Smilga:2017arl}
A.~Smilga, \emph{{Classical and quantum dynamics of higher-derivative
  systems}}, \href{http://dx.doi.org/10.1142/S0217751X17300253}{\emph{Int. J.
  Mod. Phys.} {\bf A32} (2017) 1730025},
  [\href{http://arxiv.org/abs/1710.11538}{{\tt 1710.11538}}].

\bibitem{Buchbinder:2018lbd}
I.~L. Buchbinder, E.~A. Ivanov, B.~S. Merzlikin and K.~V. Stepanyantz,
  \emph{{Gauge dependence of the one-loop divergences in $6D$, ${\cal N} =
  (1,0)$ abelian theory}},
  \href{http://dx.doi.org/10.1016/j.nuclphysb.2018.10.005}{\emph{Nucl. Phys.}
  {\bf B936} (2018) 638--660}, [\href{http://arxiv.org/abs/1808.08446}{{\tt
  1808.08446}}].

\bibitem{Bossard:2015dva}
G.~Bossard, E.~Ivanov and A.~Smilga, \emph{{Ultraviolet behavior of 6D
  supersymmetric Yang-Mills theories and harmonic superspace}},
  \href{http://dx.doi.org/10.1007/JHEP12(2015)085}{\emph{JHEP} {\bf 12} (2015)
  085}, [\href{http://arxiv.org/abs/1509.08027}{{\tt 1509.08027}}].

\bibitem{Gracey:2015xmw}
J.~A. Gracey, \emph{{Six dimensional QCD at two loops}},
  \href{http://dx.doi.org/10.1103/PhysRevD.93.025025}{\emph{Phys. Rev.} {\bf
  D93} (2016) 025025}, [\href{http://arxiv.org/abs/1512.04443}{{\tt
  1512.04443}}].

\bibitem{Gracey:2016zug}
J.~A. Gracey, \emph{{$\beta$-functions in higher dimensional field theories}},
  \href{http://dx.doi.org/10.22323/1.260.0063}{\emph{PoS} {\bf LL2016} (2016)
  063}, [\href{http://arxiv.org/abs/1610.04447}{{\tt 1610.04447}}].

\bibitem{Gusynin:1988zt}
V.~P. Gusynin, \emph{{Seeley-Gilkey Coefficients for the Fourth Order Operators
  on a Riemannian Manifold}},
  \href{http://dx.doi.org/10.1016/0550-3213(90)90233-4}{\emph{Nucl. Phys.} {\bf
  B333} (1990) 296}.

\end{thebibliography}\endgroup
\bibliographystyle{JHEP}
\end{document}